\documentclass{pasj00}

\usepackage{lineno}
\usepackage[round]{natbib}

\begin{document}
\SetRunningHead{M. Serino et al.}{Superburst with Outburst from EXO 1745-248}
\Received{2012/01/31}
\Accepted{2012/03/02}

\title{Superburst with Outburst from EXO 1745-248 in Terzan 5 with MAXI}
\author{
  Motoko \textsc{Serino},\altaffilmark{1}
  Tatehiro \textsc{Mihara}, \altaffilmark{1}
  Masaru \textsc{Matsuoka}, \altaffilmark{1}
  Satoshi \textsc{Nakahira}, \altaffilmark{1}
  Mutsumi \textsc{Sugizaki}, \altaffilmark{1}
  \\
  Yoshihiro \textsc{Ueda}, \altaffilmark{2}
  Nobuyuki \textsc{Kawai}, \altaffilmark{3}
  and
  Shiro \textsc{Ueno} \altaffilmark{4}
}
\altaffiltext{1}{MAXI team, Institute of Physical and Chemical Research (RIKEN), 2-1 Hirosawa, Wako, Saitama 351-0198}
\altaffiltext{2}{Department of Astronomy, Kyoto University, Oiwake-cho, Sakyo-ku, Kyoto 606-8502}
\altaffiltext{3}{Department of Physics, Tokyo Institute of Technology, 2-12-1 Ookayama, Meguro-ku, Tokyo 152-8551}
\altaffiltext{4}{ISS Science Project Office, Institute of Space and Astronautical Science (ISAS), Japan Aerospace Exploration Agency (JAXA), 2-1-1 Sengen, Tsukuba, Ibaraki 305-8505}
\email{motoko@crab.riken.jp}


\maketitle

\begin{abstract}

MAXI/GSC detected a superburst from EXO 1745-248 in the globular cluster 
Terzan 5 on 2011 October 24.
The GSC light curve shows an exponential decay with an 
e-folding time of 0.3 day. The spectra are consistent with the blackbody radiation, 
whose temperature is 2.2 keV and 1.2 keV at MJD 55858.56 
and 55859.20, respectively.
The fluence is $1.4 \times 10^{42}$ erg in 2-20 keV assuming 8.7 kpc 
distance.
The sphere radius of the blackbody and its luminosity are estimated to 
be 6.2 km and $1.1 \times 10^{38}$ erg s$^{-1}$, respectively,
from the spectral fitting at the flux peak.
Those e-folding time, temperature, softening, fluence, and radius are 
typical of superbursts from the low-mass X-ray binaries. 
The superburst was followed by an outburst 28 hours after the superburst 
onset.  The outburst lasted for 5 days and the fluence was $4.3 \times 
10^{42}$ erg.
The instability of 
the accretion disk caused by the superburst would be an explanation
for the outburst, whereas 
the mass accretion of the matter evaporated from surface 
of the companion star by the superburst would be another possibility.

\end{abstract}

\section{Introduction}

A thermonuclear runaway on the surface of the neutron star is 
considered as an origin of an X-ray burst. 
It normally lasts for 10-100 s.
X-ray bursts with extraordinary long durations were discovered 
from 4U 1735-44 \citep{Cornelisse2000} 
and KS 1731-260 \citep{Kuulkers2002KS},
which lasted several thousand - tens of thousand s.
It was named as a superburst.
The long duration is considered to come from the deep ignition in the carbon layer.
The fuel is mixture of carbon, hydrogen and helium 
\citep{StrohmayerBrown2002}.
It is a rare phenomenon since the recurrent time is longer than tens 
of years.
The superbursts from six sources are summarized by 
\citet{StrohmayerBildsten2006}.
The superburst was also discovered from a transient source 4U 1608-522 
\citep{2008A&A...479..177K}.
Thus, it is considered to be a common phenomenon
in X-ray bursters.

The RXTE/PCA bulge scan detected an increase of the flux from 
the globular cluster Terzan 5
on MJD 55860 ($=$ 2011 October 26) \citep{2011ATel.3714....1A}.
They mentioned that this could be a new outburst, which is a brightening of 
a transient source lasting typically some weeks to months.
The flux was 8 mCrab at MJD 55860.2 and increased to 83 and 90 mCrab
at MJD 55860.96 and 55861.03, respectively, in 2--16 keV.
\citep{2011ATel.3720....1A}.
\citet{VovkAtel} reported that INTEGRAL observed Terzan 5 region
during MJD 55859.7--55859.9, which was just before RXTE/PCA observation, 
and obtained the 5 sigma upper limits of 6 mCrab in 3--10 keV
and 11 mCrab in 20-40 keV.
MAXI had detected a brightening of Terzan 5 on MJD 55858.53
\citep{MiharaAtel}, prior to the INTEGRAL observations. 

The position of the source determined by Swift XRT is consistent
with EXO 1745-248 \citep{2011ATel.3720....1A}, 
rather than IGR J17480-2446 which had an outburst in 2010 October 
\citep{2010ATel.2919....1B,2011MNRAS.418..490C}.
This identification is confirmed by 
the following Chandra observation 
\citep{PooleyAtel} to be CXOGlb J174805.2-244647 or CX3 in \cite{Heinke2006}.
EXO 1745-248 had an outburst in 2000 
\citep{2000IAUC.7454....1M, 2005ApJ...618..883W}.
The distance to Terzan 5 was estimated to be 8.7 kpc \citep{Cohn2002}.
\citet{2011ApJ...730L..32M} found kilohertz quasi-periodic oscillations 
from this source at $\sim$ 690 Hz.
Although any orbital period of the system and the nature of the companion star
are unclear, \citet{2003ApJ...590..809H} suggested the possibility of an 
ultracompact binary system from the spectral analysis. 

We report the results of MAXI/GSC observation of the brightening episode from MJD 55858 
and the discovery of the superburst and the succeeding outburst.
We employ the distance of 8.7 kpc to Terzan 5 in this paper.

\section{Observations and Data Analysis}

MAXI/GSC 
\citep{Miharapasj}
observed Terzan 5 region with two GSC cameras (GSC-5 and GSC-B),
throughout this brightening episode.
Since GSC-B was operated with a reduced high voltage, the data 
was not included in the preliminary analysis of ATel \citep{MiharaAtel}.
In this paper, we use all the data from both cameras.

\begin{figure*}
   \begin{center}
         \FigureFile(80mm,50mm){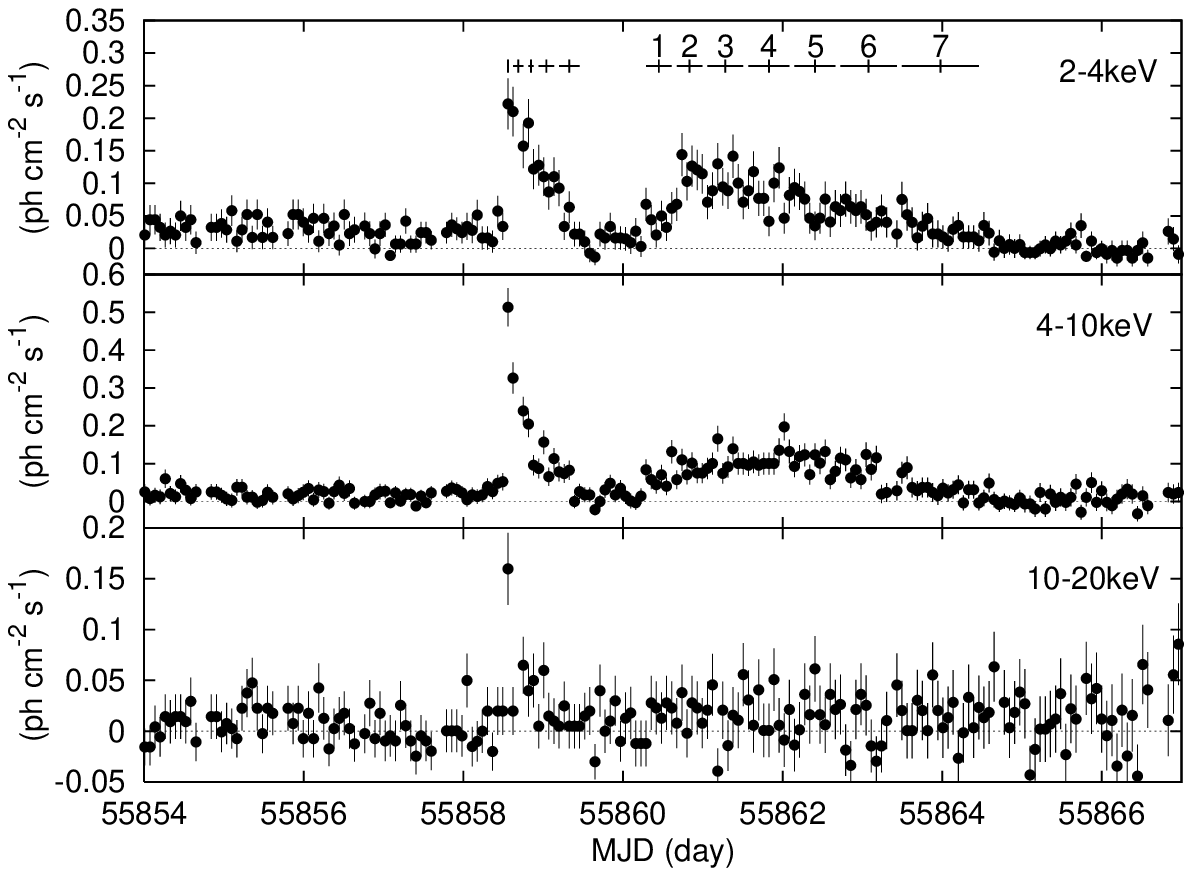}
         \FigureFile(80mm,50mm){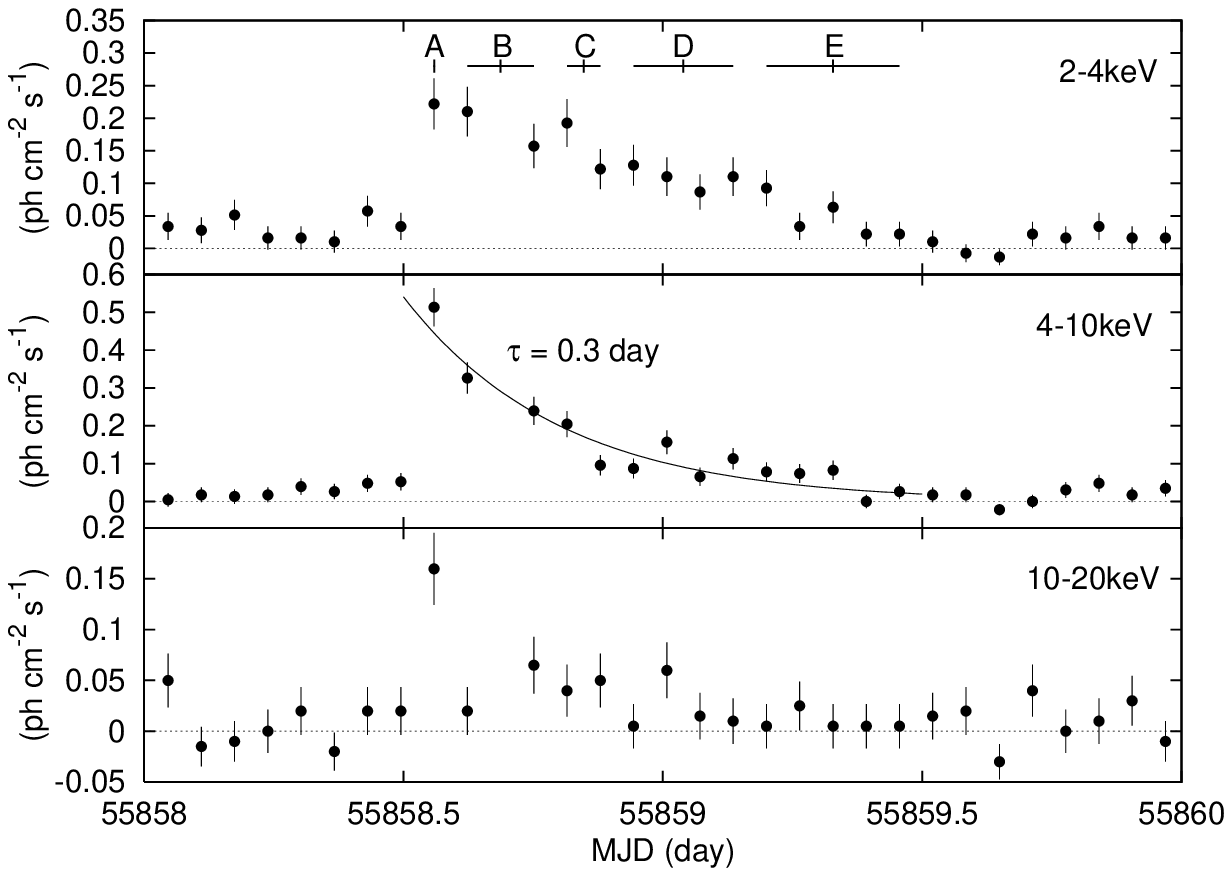}
   \end{center}
   \caption{MAXI/GSC light curves of Terzan 5 in 2--4 keV (top), 
   4--10 keV (middle), and 10--20 keV (bottom) energy bands.
   The data represent flux for each scan
   and the error bars represent the 1-$\sigma$ statistical uncertainties.
   The time is the center of the scan.
   The left panels shows the light curves between MJD 55854 (October 20)
   and MJD 55867 (November 2). The light curves between MJD 55858 
   and MJD 55860 are expanded in the right panels.
   The time intervals used for hardness plots are 
   labeled in the top panels (see figure \ref{fig:cc} and \ref{fig:hi}).
   }\label{fig:lc-burst}
\end{figure*}

A GSC camera scans a source in every 92 minutes with a typical transit time
of $\sim 60$ s \citep{2011arXiv1102.0891S}. Figure \ref{fig:lc-burst} 
shows the light curve of Terzan 5 with MAXI/GSC.  
The observed count-rates are converted to fluxes in units of photons cm$^{-2}$ s$^{-1}$
by assuming 
the Crab spectrum in each energy band.
The 1 Crab is 1.87, 1.24 and 0.40  ph cm$^{-2}$ s$^{-1}$ in 2-4, 4-10, 10-20 keV band, respectively\footnote{http://maxi.riken.jp/top/index.php?cid=36}.

\subsection{Light curves and persistent emissions}

The light curve 
shows two peaks.
The initial peak was observed only by MAXI/GSC.
The flux increased 
suddenly  at
MJD 55858.5589 from the nominal flux level of 0.02 ph cm$^{-2}$ s$^{-1}$ 
to 0.51 ph cm$^{-2}$ s$^{-1}$ (4--10 keV).
There was no flux variation during the scan ($\sim60$ s).
The flux was nominal in the 
previous scan at MJD 55858.4949, which was 92 minutes before. 
The 4-10 keV light curve showed an exponential decay after it.
The e-folding time was obtained to be 0.3 day by a fitting, 
which is typical for a superburst.
The total fluence between 55858.5 and
55859.5 is $2 \times 10^4$ ph cm$^{-2}$ in 2-20 keV band.
It corresponds to the 
total energy of 1.4 $\times 10^{42}$ erg in 2--20 keV
if the spectrum is 1.7 keV blackbody, which is an average spectrum as
shown in the section \ref{Spectral analysis}
and table \ref{tab:hardness}.
The first other observation than MAXI was done by INTEGRAL in 55859.7--55859.9.
It was after the initial component faded out, and 
the upper limit of 6 mCrab in 3--10 keV is consistent with the MAXI light curve.

After the intermission, GSC observed the re-brightening of 
the source (figure \ref{fig:lc-burst}).
This re-brightening starts at MJD 55859.9 and lasts for about 5 days.
The flux at the peak time (MJD 55862.1) was 0.13 ph cm$^{-2}$ s$^{-1}$ 
in 4--10 keV.
This light curve is consistent with the flux observed by RXTE/PCA
in 2--16 keV,
8, 83 and 90 mCrab at 55860.2, 55860.96 and 55861.03, respectively.
The total photon fluence between 55859.9 and 55864.4 is 
$6 \times 10^4$ ph cm$^{-2}$ in 2-20 keV.
Assuming the spectrum of 2.55 keV disk blackbody
\citep{2011ATel.3720....1A},
it corresponds to the total energy of
4.3 $\times 10^{42}$ erg.

Averaging over 2 years of observation, the persistent flux from the source
is 0.02 $\pm$ 0.01 ph cm$^{-2}$ s$^{-1}$ or 
16 mCrab
in 4--10 keV.  This flux may 
contain those from other X-ray sources in Terzan 5.
The maximum 4--10 keV flux from Terzan 5 observed with MAXI/GSC 
were 0.05 ph cm$^{-2}$ s$^{-1}$ or 40 mCrab in one-day average,
excluding the time interval between MJD 55478 and 55535, 
when another X-ray source in Terzan 5, IGR J17480-2446, was on outburst.

\subsection{Spectral analysis and hardness ratio}
\label{Spectral analysis}

\begin{figure}
   \begin{center}
         \FigureFile(70mm,50mm){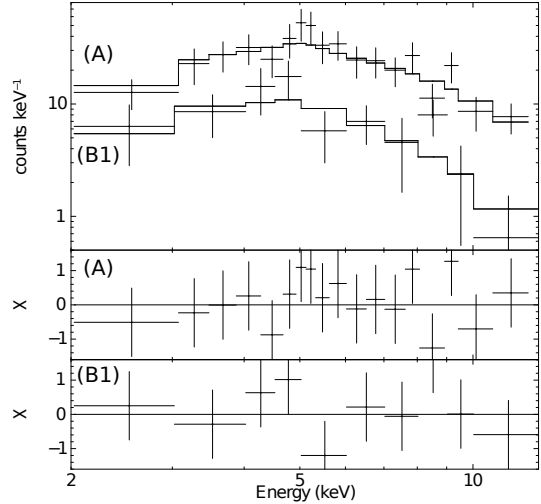}
   \end{center}
   \caption{Spectra of the time intervals
   labeled A and B1. The middle and bottom panels show the residuals
   for the spectral models summarized in table \ref{tab:hardness}.
   }\label{fig:spec}
\end{figure}

We performed time-resolved spectral analysis for the initial part.
Both a blackbody model and a power-law model ($\Gamma \sim 1$) give an acceptable fit.
Figure \ref{fig:spec} shows the spectra of the first and the second
scan with the best-fit blackbody model.
The best-fit parameters ($kT_{\rm fit}$ and $R_{\rm fit}$)  
are summarized in table \ref{tab:hardness}. 
The photo-electric absorption is not included
in the model because expected Galactic column density%
\footnote{http://heasarc.gsfc.nasa.gov/cgi-bin/Tools/w3nh/w3nh.pl}
towards Terzan 5, $N_{\rm H}=$ 5--6 $\times 10^{21}$ cm$^{-2}$,
is negligible in
fitting the GSC data.
The obtained temperature and radius are typical for a superburst.
The best-fit temperature decreased from 2.2 keV (MJD 55585.56) to 1.2 keV 
(MJD 55589.20), indicating a softening.
The luminosity in the time interval A ( = superburst peak) was 
$1.1 \times 10^{38}$ erg s$^{-1}$ if the emission is isotropic.
The temperature, softening, radius, fluence, and 
e-folding time are typical for a superburst.
Thus we conclude this is a superburst from Terzan 5.

\begin{table}
\begin{center}
\caption{Spectral parameters of the blackbody model}
\label{tab:hardness}
\begin{tabular}{ccccl}
\hline
label &
time center
           & $kT_{\rm fit}$
      & $R_{\rm fit}$\footnotemark[$\dagger$]
           & $\chi^2$ (d.o.f.) \\
 & MJD  
  & [keV] & [km] & \\
\hline
A   & 55858.56 
  & 2.2$_{-0.3}^{+0.3}$ & 6.2$_{-1.2}^{+1.3}$ &  8.9 (16) \\
B1  & 55858.62 
  & 1.7$_{-0.4}^{+0.5}$ & 6.9$_{-2.6}^{+3.8}$ &  6.5  (8) \\
B2  & 55858.75 
  & 1.7$_{-0.2}^{+0.3}$ & 7.0$_{-1.8}^{+2.3}$ &  5.7  (6) \\
C   & 55858.84 
  & 1.6$_{-0.3}^{+0.5}$ & 6.0$_{-2.5}^{+3.3}$ &  3.6  (4) \\
D+E & 55859.20 
  & 1.2$_{-0.3}^{+0.3}$ & 6.2$_{-2.4}^{+3.9}$ &  3.3  (4) \\
\hline
\multicolumn{4}{@{}l@{}}{\hbox to 0pt{\parbox{85mm}{\footnotesize
  Errors are in 90\% confidence level.
  \par\noindent
  \footnotemark[$\dagger$] 
  The distance of 8.7 kpc is assumed to calculate the radius.
}\hss}}
\end{tabular}
\end{center}
\end{table}

\begin{figure}
   \begin{center}
         \FigureFile(70mm,50mm){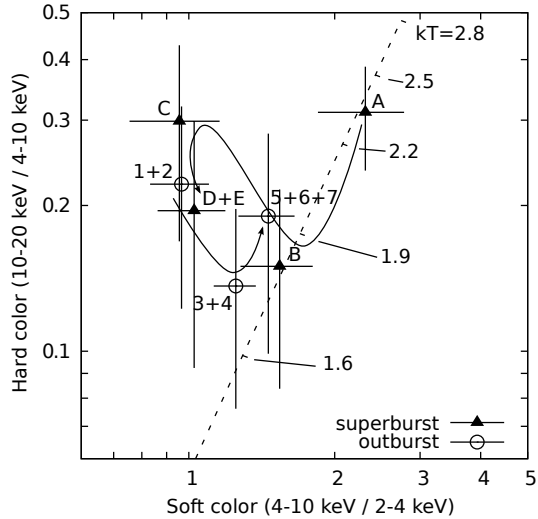}
   \end{center}
   \caption{Color-color diagram of the superburst and the outburst.
   Each point (A--E, 1--7) corresponds to the time interval shown 
   in figure \ref{fig:lc-burst}.
   The data of the superburst part are plotted with triangles and
   joined with lines (evolved from right to left).
   The data of the outburst part are plotted with circle.
   The dashed line represents the relation if the source has a blackbody spectrum.
   }\label{fig:cc}
\end{figure}

\begin{figure}
   \begin{center}
         \FigureFile(80mm,50mm){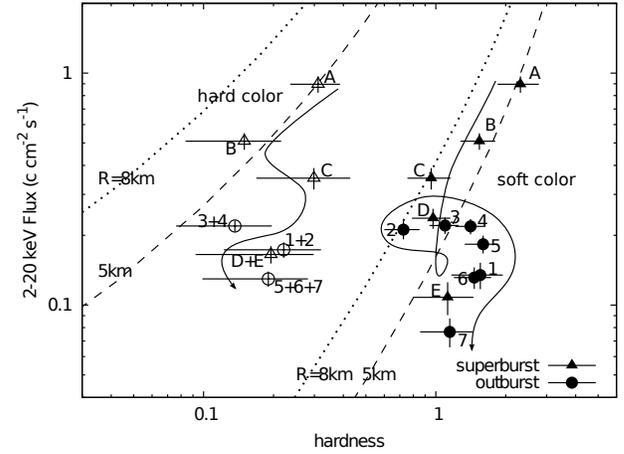}
   \end{center}
   \caption{Hardness-intensity diagram of the superburst and the outburst.
   Each point corresponds to the the data in each time interval shown 
   in the left top panel of figure \ref{fig:lc-burst}.
   The data of the superburst part are plotted with triangles and
   joined with lines.
   The data of the outburst part are plotted with circle.
   The numbers indicate the time intervals.
   The hardness of the soft and hard band are plotted with
   filled and open marks respectively.
   The relations between the hardness and the 
   flux, assuming blackbody radiation with a radius of 5 km and 
   8 km, are plotted with the dashed and dotted lines respectively.
   }\label{fig:hi}
\end{figure}

	In order to investigate the spectral evolution of the superburst
	and the following outburst, we plotted 
	color-color diagram (figure \ref{fig:cc}) and hardness-intensity
	diagram (figure \ref{fig:hi}). 
	In figure \ref{fig:cc}, the first two points of the superburst part
	agrees with the blackbody model, while the rest
	points are above the line of blackbody model.

	The curve of the soft color in figure \ref{fig:hi} shows
	a decrease of the flux by an order of magnitude along the same
        blackbody radius
	for the superburst part (triangles). On the other hand,
	there is clockwise motion in the diagram for the outburst part
	(circles). The flux stays almost constant from the interval 2 to 4, 
	while the hardness changes from 0.7 to 1.4.
	These distinctions suggest different emission
	mechanisms between the two parts.
	In fact \citet{2011ATel.3720....1A} reported that
	the spectrum of the latter part
	is represented by a model consisting of  diskbb of $kT_{\rm in}$ = 2.55 keV
	powerlaw of photon index = 2.25, Gaussian iron line (6.5 keV),
	and absorption (column density $= 1.2 \times 10^{22}$ cm$^{-2}$, fixed),
	suggesting the emission of the outburst comes from
	the accretion disk rather than neutron star surface.

\section{Discussion}

\subsection{Disk instability caused by the superburst}
\label{disk instability}
	How did the superburst trigger the outburst ?
	\cite{2009A&A...503..889K} discussed a possibility that a normal
	X-ray burst induced the instability of the accretion disk and triggered 
	outburst of Cen X-4.
	In the case of Cen X-4, they mentioned the duration of the irradiation 
	of 10 minutes is not sufficient to modify the thermal structure of the
	disk. 
	Using the disk viscosity parameter $\alpha$,
	the characteristic thermal time scale $t_{\rm th}$ is
	\begin{eqnarray}
		t_{\rm th} &=& \alpha^{-1} \Omega_{\rm K}^{-1} 
		\nonumber \\
		          &=& \alpha^{-1} (G M_1 / R^3)^{-1/2} \, ,
	\end{eqnarray}
	where $\Omega_{\rm K}$, $G$, and $M_1$ are the Keplarian angular 
	velocity gravitational constant and the mass of the neutron star.
	It is $\sim$ 4 min when the radius $R = 10^9$ cm, $\alpha \sim 0.01$, 
	and $M_1=1.4 M_{\odot}$ are assumed.

	In the case of the superburst, the duration is several hours and 
	it is much longer than the thermal time scale, even at the
	outer boundary of the disk. 
	Therefore the disk-instability scenario
	is a possible model to explain the outburst.

\subsection{Evaporation of the companion-star surface}
	Another explanation of the outburst is mass accretion of the matter 
	evaporated from surface of the companion star by the superburst.
	We define $d_1$ and $d_2$ as the distance from the inner Lagrangian 
	point $L_1$ to the neutron star and to the companion star, respectively.
	As typical to the low-mass X-ray binaries, the companion
	low-mass star is assumed to fill the Roche lobe.
	Therefore, the radius of the companion star is $R_2 \simeq d_2$.

	The total radiation energy of the superburst $E_{\rm s}$
	heats up the companion star, whose solid angle from the neutron star is
	$\Omega$, with the efficiency $\eta$, which is defined as the 
	efficiency of the irradiated energy to the kinetic 
	energy of the escaping matter.
	Then, the matter on the surface with mass $m$ obtains the kinetic energy and escapes
	from the surface with the velocity $v$. The energy conservation gives
	\begin{equation}
		\label{eq:escape}
		v = \sqrt{2 \eta  \, E_{\rm s} \Omega/ (4 \pi m)} \, .
	\end{equation}
	The solid angle $\Omega$ is written as
	\begin{equation}
		\Omega = \frac{\pi {R_2}^2}{(d_1+R_2)^2} \, .
	\end{equation}
	The mass of the companion star is assumed to be those of the typical X-ray bursters,
	$0.04 M_{\odot} \leq M_2 \leq 0.4 M_{\odot}$ 
	\citep{1993SSRv...62..223L}. 
        Since the mass-to-radius ratio is almost 
	constant among main sequence stars,
        the radius of the companion star with the mass in this range becomes $0.04 R_{\odot} \leq R_2 \leq 0.4 R_{\odot}$.
        Then, the ratio $R_2 / d_1 $ is calculated to be  $0.24 < R_2 / d_1 < 0.59$,
        assuming the mass of the neutron star $M_1$ of 1.4 $M_{\odot}$.
	Thus $\Omega$ is constrained to be
	\begin{equation}
	0.12 \ltsim \Omega \ltsim 0.44 \, . 
	\label{eq:omega}
	\end{equation}

	Supposing that the escaped matter from the companion star surface 
        accretes onto the neutron star,
	the velocity of the evaporated matter should be faster than
	the escape velocity of the companion star to the Lagrangian point
	$L_1$, but should not exceed that of the binary system.
	The escape velocity of the companion star to infinity
	is written as $\sqrt{2GM_2/R_2}$.
	Let us express the escape velocity to $L_1$ with a parameter $\epsilon$,
	as $\sqrt{2 \epsilon GM_2/R_2}$.
	Under the assumption of the constant mass-to-radius ratio,
	it is written as $\sqrt{2 \epsilon GM_\odot/R_\odot}$.
	The escape velocity is expressed as
	$\sim 6 \times 10^7 \epsilon^{1/2}$ cm s$^{-1}$. 
	When the companion star completely fills the Roche lobe, $\epsilon \sim 0$.
	When the escape velocity is equal to
	the thermal velocity of 3000~K atmosphere,
	which is typical of a low-mass star, $\epsilon \sim 0.0001$. 

	The escape velocity out of the system is estimated as 
	$\sim 2 \times 10^8$ cm s$^{-1}$ with the mass of $1.4 M_\odot$ and 
	typical size of the low-mass X-ray binary $10^{10}$ cm.
	Therefore, the velocity of the evaporated matter must be
	\begin{equation}
		\label{eq:vcond}
		6 \times 10^7 \epsilon^{1/2} \leq v \leq 2 \times 10^8 
		\hspace{1em} {\rm cm ~ s}^{-1} \, .
	\end{equation}

        The total accreted mass in the outburst $m$ can be calculated from the 
	fluence of the outburst (4.3 $\times 10^{42}$ erg)
	and the energy release of accreting matter per unit mass,
	$G M_1 / R_1 \sim 2 \times 10^{20}$ erg g$^{-1}$,
	where $R_1$ is the radius of the neutron star.
	They lead to $m = 2.2 \times 10^{22}$ g.
        The mass column density of the evaporated gas on the companion hemisphere surface becomes 
        $m/(2\pi R_2^2) = 4.5 \sim 450 $ g cm$^{-2}$ for $R_2 = (0.04 \sim 0.4) R_\odot$.
	Substituting the accreted mass $m$ and $E_{\rm s} = 1.2\times 10^{42}$ erg into
	equations (\ref{eq:escape}) and (\ref{eq:vcond}),
	the condition to $\eta \Omega$ is obtained as 
	$3.9 \times 10^{-4} \epsilon \ltsim \eta \Omega 
	\ltsim 3.9 \times 10^{-3}$.
	Considering equation (\ref{eq:omega}), the constraint on $\eta$ 
	is obtained as $9 \times 10^{-4} \epsilon \ltsim \eta \ltsim 0.03$.

	The course of surface-matter evaporation by X-ray 
	irradiation contains many elementary processes
	such as 
	Compton scattering by electrons (hydrogen plasma),
	photo-electric absorption by the metal elements,
	thermal conduction from electrons to protons, and
	escape probability of the proton from the reaction point to outer space
	through atmosphere whose density varies exponentially with height.
	Precise theoretical calculation would be necessary to evaluate $\eta$,
	which is beyond the scope of this paper.

\subsection{Feedback by the mass evaporation}

	The evaporation model contains a feedback problem.
	If the superburst X-ray induced the succeeding outburst,
	the succeeding outburst itself would also induce another outburst since the flux of the 
	initial superburst and the secondary outburst is comparable.
	Then the outburst is supposed to repeat forever by this positive feedback.
	The fact is that the secondary outburst occurred only once and stopped.

	The energy spectrum is similar in the superburst and the outburst and 
	would not cause a difference.
	Both have the temperature of $\sim 2$ keV. In fact, the initial spike of 
	the superburst was hard and had the only positively-detected data point in the 10-20 keV band 
	as shown in figure \ref{fig:lc-burst}. However, this spike contains only a small 
	fraction of the fluence of the superburst.
	
	A possible reason might be formation of the accretion disk.
	If there was no, or only a small, accretion disk when the superburst occurred,
	the superburst could irradiate the whole companion surface.
	However, once the accretion disk is formed by the mass transfer, 
	a significant portion of the X-ray emission
	towards the companion star is shadowed by the disk. 
	It reduces the evaporation of the companion surface and prevents the feedback.

	Another possible reason might be the duration of superburst heating.
	The superburst lasted for 0.3 day while the accretion lasted for 5 
	days. The flux of the outburst was only 0.2 times of that of the superburst.
	There might be a threshold to induce effective mass evaporation and accretion between
	$1.4 \times 10^{42} \,{\rm erg} / 0.3 \,{\rm day} = 5.4 \times 10^{37}$ 
	erg s$^{-1}$ and $4.3 \times 10^{42} \,{\rm erg} / 5 \,{\rm day} = 
	1.0 \times 10^{37}$ erg s$^{-1}$.
	The efficiency $\eta$ may depend on the flux.

	If the system of EXO 1745-248 is typical for LMXBs,
	the interpretation above leads to the idea that a bright enough 
	($>5.4 \times 10^{37}$ erg s$^{-1}$ or more if the shadow is considered) 
        outburst of LMXB induces the following 
	mass accretion by the positive feedback.
        The outburst does not stop 
	immediately ($<$1d), but lasts for a long period ($\gtsim$ 1 week).
	It may be difficult to distinguish whether 
        the continuing outburst is due to the
	positive feedback by the evaporation, or just due to the increased stellar 
	outflow activity. 
        However, this is at least consistent with the observational 
	facts \citep[e.g. review by][]{2004NuPhS.132..486I}
	that once the outbursts in LMXB start they last for a while 
	(some weeks to months).

\subsection{The recurrence time of the superburst}
	From the released energy by the superburst and the persistent flux,
	the recurrence time can be estimated.
	Using the fluence of the superburst of 1.4 $\times 10^{42}$ erg 
	and hydrogen burning energy of 6 $\times 10^{18}$ erg g$^{-1}$, 
	the mass of the hydrogen fuel is 2.3 $\times 10^{23}$ g. 
	Using the average flux of 16 mCrab 
	(2.7 $\times 10^{36}$ erg s$^{-1} =$ 1.4 $\times 10^{16}$ g s$^{-1}$)
	for the persistent accretion, 
	the recurrence time is calculated to be 190 days.  
	Since the energy release of the carbon burning is  $\sim 10^{18}$ 
	erg g$^{-1}$ \citep{StrohmayerBrown2002}, the recurrence time
	should be longer to be 1140 days in the carbon burning case.

\section{Conclusion}
MAXI/GSC detected a superburst from EXO 1745-248 in Terzan 5.
The light curve obtained by MAXI shows an exponential decay with an 
e-folding time of 0.3 day. From the spectral analysis of each scan,
we found a softening of the blackbody temperature, from 2.2 keV 
(MJD 55858.56) to 1.2 keV (MJD 55859.20).
The sphere radius of the blackbody and its luminosity are estimated to 
be 6.2 km and $1.1 \times 10^{38}$ erg s$^{-1}$, respectively, 
at the flux peak assuming 8.7 kpc distance.
The total fluence of the superburst between 55858.5 and
55859.5 is $1.4 \times 10^{42}$ erg in 2-20 keV.
Those e-folding time, temperature, softening, fluence, and radius are 
typical of superbursts from low-mass X-ray binaries. 

An outburst followed the superburst
28 hours after the superburst onset.
The outburst lasted for 5 days and the fluence was $4.3 \times 10^{42}$ 
erg.
We showed that the timescale and the total energy of the
outburst could be consistent with the assumption that 
the matter evaporated from surface of the companion star 
by the superburst accreted onto the neutron star,
while there is a possibility that the instability of the accretion disk
caused by the superburst is responsible for the outburst.

\bigskip

This research was partially supported by the Ministry of Education, 
Culture, Sports, Science and Technology (MEXT), Grant-in-Aid for
Science Research 20244015.

\end{document}